\documentclass[twocolumn,showpacs,preprintnumbers,amsmath,amssymb]{revtex4}

\usepackage{graphicx}
\usepackage{dcolumn}
\usepackage{bm}

\begin{document}

\title{Continuous-time quantum walks on one-dimension regular networks}
\author{Xin-Ping Xu$^{1,2}$}
\email{xuxp@mail.ihep.ac.cn}
\affiliation{%
$^1$Institute of Particle Physics, HuaZhong Normal University, Wuhan
430079, China \\
$^2$Institute of High Energy Physics, Chinese Academy of Science,
Beijing 100049, China
}%

\date{\today}

\begin{abstract}
In this paper, we consider continuous-time quantum walks (CTQWs) on
one-dimension ring lattice of $N$ nodes in which every node is
connected to its $2m$ nearest neighbors ($m$ on either side).  In
the framework of Bloch function ansatz, we calculate the spacetime
transition probabilities between two nodes of the lattice. We find
that the transport of CTQWs between two different nodes is faster
than that of the classical continuous-time random walk (CTRWs). The
transport speed, which is defined by the ratio of the shortest path
length and propagating time, increases with the connectivity
parameter $m$ for both the CTQWs and CTRWs. For fixed parameter $m$,
the transport of CTRWs gets slow with the increase of the shortest
distance while the transport (speed) of CTQWs turns out to be a
constant value. In the long time limit, depending on the network
size $N$ and connectivity parameter $m$, the limiting probability
distributions of CTQWs show various patterns. When the network size
$N$ is an even number, the probability of being at the original node
differs from that of being at the opposite node, which also depends
on the precise value of parameter $m$.
\end{abstract}
\maketitle
\section{Introduction}
Quantum walks have important applications in various fields of
solid-state physics, polymer chemistry, biology, astronomy,
mathematics and computer science~\cite{rn32,rn33,rn1,rn2,rn3,rn4}. A
quantum random walk (QRW) is a natural extension to the quantum
world of the ubiquitous classical random walk. It was first
introduced in \cite{rn5} and extensively investigated recently in
connection with possible applications to quantum algorithms
\cite{rn6}. The behavior of quantum walks differs from that of the
classical random walks in several striking ways, due to the fact
that quantum walks exhibit interference patterns whereas the
classical random walks do not. For instance, the mixing times,
hitting times and exit probabilities of quantum walks can differ
significantly from analogously defined random walks
\cite{rn7,rn8,rn9}. In recent years, two types of quantum walks
exist in the literature: the discrete-time quantum coined walks and
continuous-time quantum walks \cite{rn10,rn11}. Although both the
two types of quantum walks have efficient quantum algorithms with
respect to their classical counterparts, quantum walks show some
advantages in dealing with decoherence processes compared to the
discrete-time quantum algorithms, which are very sensitive to
environmental quantum noise \cite{rn12}.

Here, we focus on continuous-time quantum walks (CTQWs). Most of
previous studies consider CTQWs on simple structures, such as, the
line \cite{rn13,rn14}, cycle \cite{rn15,rn16}, hypercube
\cite{rn17}, Cayley tree \cite{rn18}, dendrimers \cite{rn19} and
other regular networks with simple topology. Although CTQWs have
received much attention and there has been some work about CTQWs on
general graphs, many questions about CTQWs appear to be quite
difficult to answer at the present time. For simple structures these
quantum walks are analytical solvable and directly related to the
well-known problems in solid state physics. Recently, Oliver
M\"{u}lken et al have studied the spacetime structures of CTQWs on
one-dimensional and two-dimensional lattices with periodic boundary
conditions \cite{rn20,rn21}. The topology of the lattices they
considered is oversimplified, i.e., each node is only connected to
its two nearest neighbors. For regular graphs with symmetrical
structure, the dynamics of the quantum transport is determined by
the topology of the network. To this end, it is natural to consider
quantum transport on general lattices with more connectivity.

In this paper, we study CTQWs on one-dimension ring lattice of $N$
nodes in which every node is connected to its $2m$ nearest neighbors
($m$ on either side). This generalized regular network has broad
applications in various coupled dynamical systems, including
biological oscillators \cite{rn22}, Josephson junction arrays
\cite{rn23}, neural networks \cite{rn24}, synchronization
\cite{rn25}, small-world networks \cite{rn26} and many other
self-organizing systems. We analyze quantum walks on such general
network with periodic boundary conditions using the Bloch function
approach~\cite{rn27}, which is commonly used in solid state physics.
We derive analytical expressions for the transition probabilities
between two nodes of the networks, and compare them with the results
of CTRWs.

The paper is structured as follows: In Sec.
\uppercase\expandafter{\romannumeral 2} we review the properties of
CTQWs presented in Ref.\cite{rn28} and give the exact solutions to
the transition probabilities on the general ring network. Section
\uppercase\expandafter{\romannumeral 3} presents the time evolution
of the probabilities. In Sec. \uppercase\expandafter{\romannumeral
4}, we consider the distributions of long time limiting
probabilities. Conclusions and discussions are given in the last
part, Sec. \uppercase\expandafter{\romannumeral 5}.
\section{Continuous-time quantum walks}
Keeping in line with previous results on quantum walks, we study
continuous-time quantum walk on networks and compare the results
with the classical counterparts.
\subsection{Continuous-time quantum walks on general networks}
We consider a walk on a general graph, which is a collection of
connected nodes and simple links without weight and directions. The
topology of such simple graphs can be described by the corresponding
Laplace matrix $A$. The nondiagonal elements $A_{ij}$ equal to $-1$
if nodes $i$ and $j$ are connected and $0$ otherwise. The diagonal
elements $A_{ii}$ equal the number of total links connected to node
$i$, i.e., $A_{ii}$ equals to the degree of node $i$. Classically,
the evolution of continuous-time random walk is governed by the
master equation \cite{rn1}
\begin{equation}\label{eq1}
\frac{dp_{k,j}}{dt}=\sum_{l}T_{kl}p_{l,j}(t),
\end{equation}
Where $p_{k,j}(t)$ is the conditional probability to find the CTRW
at time $t$ at node $k$ when starting at node $j$. Matrix $T$ is the
transfer matrix of the walk, and relates to the Laplace matrix by
$T= -\gamma A$. Here, for the sake of simplicity, we assume the
transmission rate $\gamma$ for all connections to be equal. Then the
solution of the above equation is
\begin{equation}\label{eq2}
p_{k,j}(t)=<k|e^{tT}|j>.
\end{equation}
Quantum mechanically, the dynamical evolution equation of
continuous-time quantum walks is obtained by replacing the
Hamiltonian of the system by the classical transfer matrix, $H=-T$
\cite{rn7,rn8}. The states $|j>$ endowed with the nodes $j$ of the
network form a complete, ortho-normalised basis set, which span the
whole accessible Hilbert space, i.e., $\sum_k |k><k|=1$,
$<k|j>=\delta_{kj}$. The time evolution of state $|j>$ is given by
the Schrodinger Equation (SE)
\begin{equation}\label{eq3}
i\frac{d|j>}{dt}=H|j>,
\end{equation}
Where the mass $m\equiv 1$ and $\hbar \equiv 1$ is assumed in the
above equation. Starting at time $t_0$ from the state $|j>$, the
evolution equation of the state $|j>$ is $|j,t>=U(t,t_0)|j>$, where
$U(t,t_0)=e^{-iH(t-t_0)}$ is the quantum mechanical time evolution
operator. The transition amplitude $\alpha_{k,j}(t)$ from state
$|j>$ at time $0$ to state $|k>$ at time $t$ is
\begin{equation}\label{eq4}
\alpha_{k,j}(t)=<k|e^{-itH}|j>,
\end{equation}
Combining Eq. (\ref{eq3}), we have
\begin{equation}\label{eq5}
i\frac{d\alpha_{k,j}}{dt}=\sum_{l}H_{kl}\alpha_{l,j}(t),
\end{equation}
We note that the different normalization for CTRWs and CTQWs. For
CTRWs, $\sum_k p_{k,j}=1$ and quantum mechanically $\sum_k
|\alpha_{k,j}|^2=1$ holds.

 To get the exact solution of Eqs. (\ref{eq1}) and (\ref{eq5}), all the eigenvalues and eigenvectors of the transfer
operator and Hamiltonian are required. We use $E_n$ to represent the
$n$th eigenvalue of $A$ and denote the orthonormalized eigenstate of
Hamiltonian by $|q_n>$, such that $\sum_n|q_n><q_n|=1$. The
classical transition probability between two nodes is given by
\begin{equation}\label{eq6}
p_{k,j}(t)=\sum_n e^{-\gamma tE_n}<k|q_n><q_n|j>,
\end{equation}
And the quantum mechanical transition probability between $k$ and
$j$ is
\begin{equation}\label{eq7}
\begin{array}{ll}
&\pi_{k,j}(t)=|\alpha_{k,j}(t)|^2\\
&=\sum_{n,l} e^{-i\gamma t(E_n-E_l)}<k|q_n><q_n|j><k|q_l><q_l|j>.
\end{array}
\end{equation}
For finite networks, $\pi_{k,j}(t)$ do not decay ad infinitum but at
some time fluctuates about a constant value. This value is
determined by the long time average of $\pi_{k,j}(t)$
\begin{equation}\label{eq8}
\begin{array}{ll}
\chi_{k,j}&=\lim_{T\rightarrow \infty}\frac{1}{T}\int_0^T
\pi_{k,j}(t)dt\\
&=\sum_{n,l}<k|q_n><q_n|j><k|q_l><q_l|j> \\
&\  \ \  \ \times\lim_{T\rightarrow \infty}\frac{1}{T}\int_0^T e^{-i\gamma t(E_n-E_l)}dt\\
&=\sum_{n,l}\delta_{E_n,E_l}<k|q_n><q_n|j><k|q_l><q_l|j>.
\end{array}
\end{equation}
\subsection{Continuous-time quantum walks on 1D ring lattice and Bloch ansatz solutions}
In the subsequent calculation, we restrict our attention on CTQWs on
the general one-dimension ring lattices with periodic boundary
conditions. The network organizes in a very regular manner, i.e.,
each node of the lattice is connected to its $2m$ nearest neighbors
($m$ on either side), thus the Laplace matrix $A$ takes the
following form,
\begin{equation}\label{eq9}
 A_{ij}=\left\{
\begin{array}{ll}
2m,   & {\rm if} \ i=j,\\
-1,   & {\rm if} \ i=j\pm z, z\in [1,m] \\
0,    & Otherwise.
\end{array}
\right.
\end{equation}
The Hamiltonian of the system is given by $H=\gamma A$. For
simplicity of analytical treatment, we set $\gamma=1$ in further
calculations. The Hamiltonian acting on the state $|j>$ can be
written as
\begin{equation}\label{eq10}
H|j>=(2m+1)|j>-\sum_{z=-m}^m|j+z>, \ z \in \ Integers.
\end{equation}
The above Equation is the discrete version of the Hamiltonian for a
free particle moving on the lattice. Using the Bloch function
approach \cite{rn27} for the periodic system in solid state physics,
the time independent SE reads
\begin{equation}\label{eq11}
H|\psi_n>=E_n|\psi_n>.
\end{equation}
The Bloch states $|\psi_n>$ can be expanded as a linear combination
of the states $|j>$ localized at node $j$,
\begin{equation}\label{eq12}
|\psi_n>=\frac{1}{\sqrt{N}}\sum_{j=1}^N e^{-i\theta_n j}|j>.
\end{equation}
Substituting Eqs. (\ref{eq10}) and (\ref{eq12}) into Eq.
(\ref{eq11}), we obtain the eigenvalues (or energy) of the system,
\begin{equation}\label{eq13}
E_n=2m-2\sum_{j=1}^m \cos(j\theta_n)
\end{equation}
The periodic boundary condition for the network requires that the
projection of  on the state $|N+1>$ equals to that on the state
$|1>$, thus $\theta_n=2n\pi/N$ with $n$ integer and $n\in[0, N)$.
Replacing $|q_n>$ by the Bloch states $|\psi_n>$ in Eqs.
(\ref{eq6}), (\ref{eq7}), we can get the classical and quantum
transition probability
\begin{equation}\label{eq14}
p_{k,j}(t)=\frac{1}{N}\sum_n e^{-tE_n}e^{-i(k-j)\frac{2n\pi}{N}},
\end{equation}
\begin{equation}\label{eq15}
\begin{array}{ll}
\pi_{k,j}(t) &=|\alpha_{k,j}(t)|^2  \\
&=\frac{1}{N^2}\sum_{n,l} e^{-it(E_n-E_l)}e^{-i(k-j)(n-l)},
\end{array}
\end{equation}
For infinite networks, i.e., $N\rightarrow\infty$, Eqs. (\ref{eq14})
and (\ref{eq15}) translates to
\begin{equation}\label{eq16}
\lim_{N\rightarrow \infty}
p_{k,j}(t)=\frac{e^{-2mt}}{2\pi}\int_{-\pi}^{\pi} e^{-i\theta
(k-j)}e^{2t\sum_{j=1}^m\cos j\theta}d\theta,
\end{equation}
\begin{equation}\label{eq17}
\lim_{N\rightarrow \infty}
\pi_{k,j}(t)=|\frac{1}{2\pi}\int_{-\pi}^{\pi} e^{-i\theta
(k-j)}e^{2it\sum_{j=1}^m\cos j\theta}d\theta|^2,
\end{equation}
Particularly, when $m=1$, the network corresponds to a cycle graph
where each node has exact two nearest neighbors. The limiting
transition probability can be rewritten as $\lim_{N\rightarrow
\infty} p_{k,j}(t)=e^{-2mt}\ BesselJ(k-j,2t)$ and
$\lim_{N\rightarrow \infty} \pi_{k,j}(t)=[BesselJ(k-j,2t)]^2$, where
$BesselJ$ is the Bessel function of the first kind \cite{rn29}. This
is consistent with the result in Ref. \cite{rn21}. The difference
between finite and infinite network is that, for infinite networks
the interference of quantum transport is weak compared to finite
networks. For larger value of $m$, the above analytical expression
could not be further simplified. We can calculate the transition
probabilities straightly using the integrations for the infinite
networks. We will show that there is some difference of the
transition probabilities between finite and infinite networks at
long time scale.

Finally, the long time averaged probability between two nodes yields
\begin{equation}\label{eq18}
\chi_{k,j}=\frac{1}{N^2}\sum_{n,l}\delta_{E_n,E_l}e^{-i(k-j)(n-l)}.
\end{equation}
  Interestingly, the long time averaged probability is related to
the spectral of the networks, in contrast to the classical transport
where there is a uniform probability ($1/N$) to find the walker at
every node. The time limiting probabilities depend on the
degeneracies of the eigenvalues, which result in odd, unexpected
patterns of limiting probability distributions.
\section{Time evolution of the probabilities}
In this section, we analyze the time dependent probabilities of the
theoretical calculations. The numerical determination of the
eigenvalues, eigenvectors and integration is done using the software
Mathematica. Specifically, we perform our calculations on infinite
and finite ($N=100$) networks with different connectivity $m$.
\begin{figure}
\scalebox{0.8}[0.8]{\includegraphics{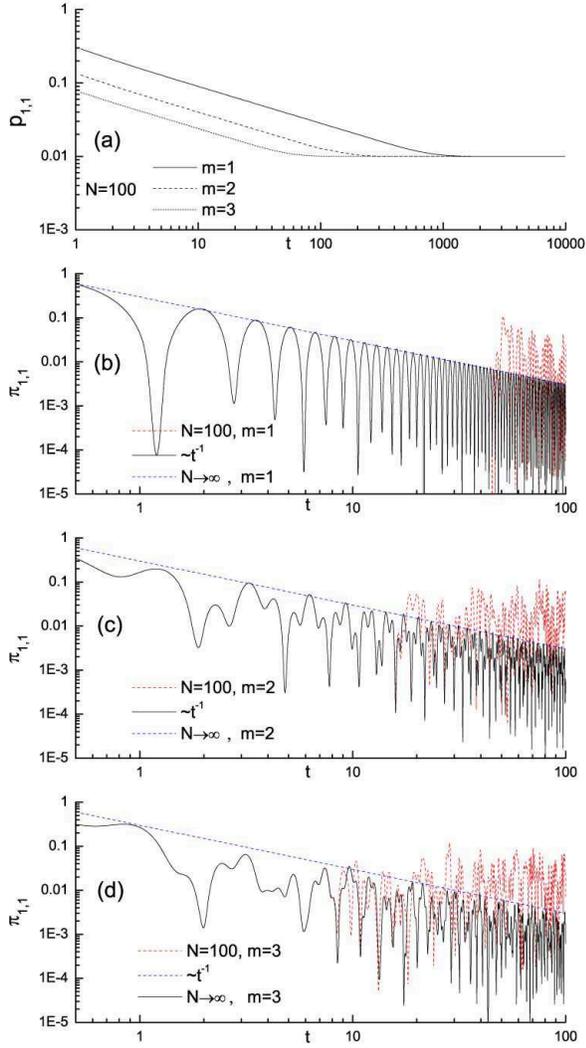}} \caption{(Color
online) Evolution of the probability of being at the initial node
$1$. (a)Classical return probability $p_{1,1}$ on networks of
$N=100$ with different values of $m$. $p_{1,1}$ approaches the
equipartitioned probability $1/N$ quickly on networks with high
connectivity. (b), (c) and (d) show the evolution of quantum
mechanical return probabilities $\pi_{1,1}$ with $m=1$, $m=2$ and
$m=3$, respectively. The dashed curves are results on network of
size $N=100$ according to Eq.(\ref{eq20}), the solid curves are the
corresponding results on infinite networks according to
Eq.(\ref{eq17}). The dashed lines show the scaling behavior
$\pi_{1,1}\sim t^{-1}$. \label{fg1}}
\end{figure}
\begin{figure}
\scalebox{0.8}[0.8]{\includegraphics{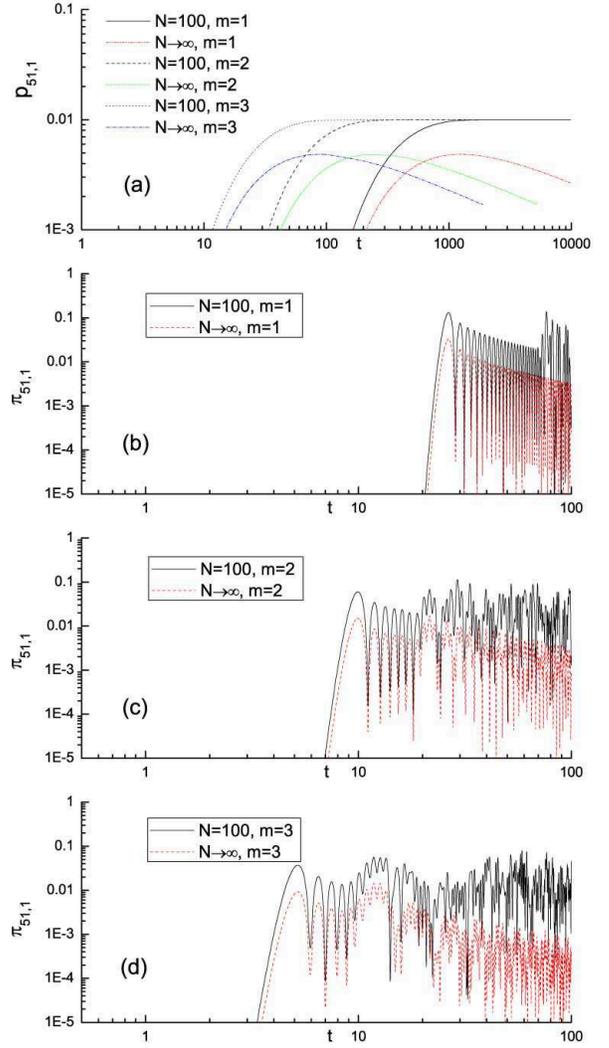}} \caption{(Color
online) Evolution of the probability of finding the walker at the
opposite node $51$ when the initial node is $1$. (a) The classical
transition probability $p_{51,1}$ for infinite networks and finite
network of $N=100$ with different parameter $m$. We can see that,
the probability of infinite network with large connectivity reaches
its maximum quicker than that of infinite network with small
connectivity. (b), (c) and (d) are the quantum mechanical transition
probabilities $\pi_{51,1}$ for $m=1$, $m=2$ and $m=3$. The solid
curves are the results on infinite networks, the dashed curves are
the results on finite networks of $N=100$. \label{fg2}}
\end{figure}
\subsection{Return probabilities}
The probability to be still or again at the initial node is a good
measure to quantify the efficiency of the transport \cite{rn30}.
Classically, according to Eq. (\ref{eq14}), the probability of being
at the original node $j$ is
\begin{equation}\label{eq19}
        p_{j,j}(t)=\frac{1}{N}\sum_n e^{-tE_n},
\end{equation}
Which only depends on the eigenvalues. The quantum-mechanical
probability of finding the walker at the initial node is given by
Eq. (\ref{eq15}),
\begin{equation}\label{eq20}
         \pi_{j,j}(t)=\frac{1}{N^2}\sum_{n,l}e^{-it(E_n-E_l)},
\end{equation}
Which also dependents on the eigenvalues of the system. The return
probability is independent on the position of the initial excitation
nodes because of the symmetry of network topology. Analogously,
employing the relation of $k=j$, we can calculate the return
probabilities on infinite networks according to Eqs (\ref{eq16}) and
(\ref{eq17}).
\begin{figure*}
\scalebox{1.0}[0.9]{\includegraphics{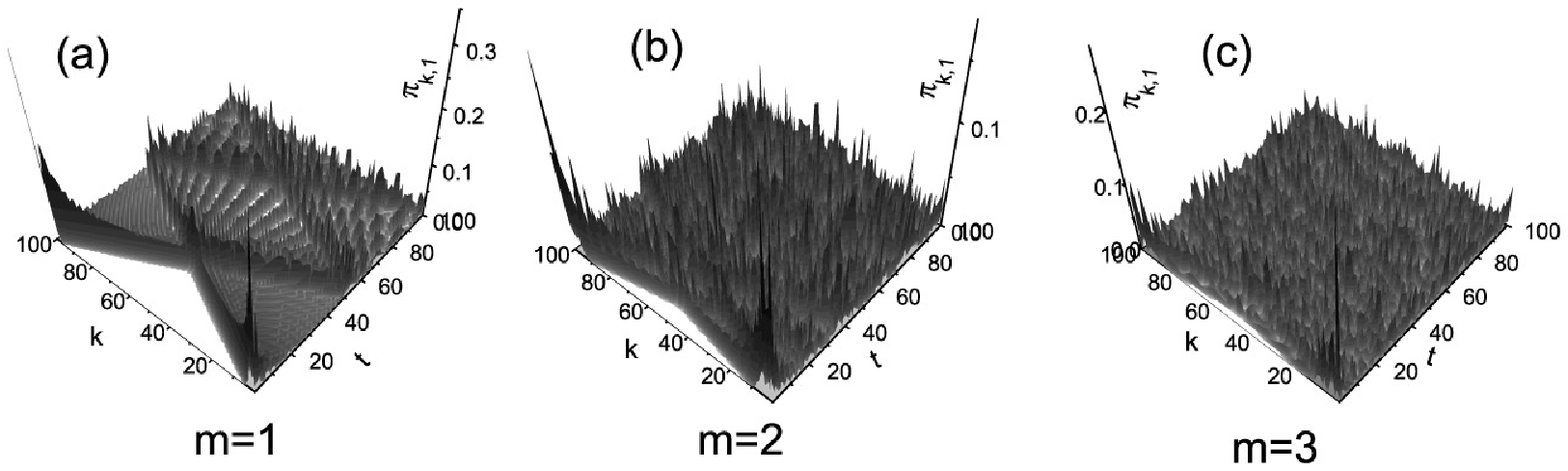}}
 \caption{(Color online)
 Development of transition probabilities $\pi_{k,1}(t)$ for CTQWs on network of
 $N=100$ with parameter $m=1$ (a), $m=2$ (b) and $m=3$ (c). The
 initial exciton starts at node $1$.
\label{fg3}}
\end{figure*}

   Fig.\ref{fg1} shows the return probabilities for CTRWs and CTQWs.
Consider a CTRW on a network of size $N=100$ and assume the initial
excitation starts at node $1$. Fig.\ref{fg1} (a) depicts the
temporal behavior of return probability with different values of
$m$. There is a power law decay ($p\sim t^{-0.5}$) at the beginning
of the transport, but after some time $p$ reaches a constant value.
This time is determined by the time when $p_{1,1}$ reaches the
equipartitioned probability $1/N$. The time becomes small when the
parameter $m$ increases, this indicates that it takes less time for
the return probability to reach the equipartitioned probability on
networks with high connectivity. Fig.\ref{fg1} (b), (c) and (d)
shows the quantum mechanical return probabilities for $m=1$, $m=2$,
and $m=3$, respectively. The dashed curves show the results on
network of $N=100$ and the black solid curves show the results on
infinite network according to Eq. (\ref{eq17}). The dashed lines
indicate the scaling behavior $\pi_{1,1}\sim t^{-1}$. We note that
the return probabilities of finite and infinite networks agree with
each other in small time scales. At later times waves propagating on
the finite networks start to interferer, this leads to the
probabilities differ and the deviation happens at earlier times on
highly connected networks (with larger value of $m$). Furthermore,
the return probabilities oscillate frequently on highly connected
networks and there are more peaks compared to networks with small
value of $m$. Such a behavior may be attributed to the fact that the
interferences on networks with high connectivity are stronger than
on those with small connectivity.
\subsection{Transition probabilities and transport velocity}
The transition probabilities between two different nodes provide us
more information about the transport process over the whole network.
For a finite network of $N=100$, we consider the probability of
finding the walker at the opposite node. Fig.~\ref{fg2} shows the
transition probabilities for CTRWs and CTQWs. In Figure~\ref{fg2},
(a) shows the classical transition probabilities $p_{51,1}$ on
infinite and finite networks of size $N=100$ with different values
of $m$. As we can see, the transition probabilities on finite
networks with more connectivity approach to the equipartitioned
probability $1/N$ quicker than those on network with less
connectivity. For infinite network, the transition probabilities
increase with time in the first period, and then reach the maxima
and decrease in the large scale time. Quantum mechanically, the
transition probabilities for $m=1$, $m=2$, and $m=3$ are shown in
Fig.~\ref{fg2} (b), (c) and (d). The dashed curves are the results
for network of $N=100$, solid curves are the corresponding results
for infinite networks. The transition probabilities on infinite
networks are smaller than those on finite networks at the same time.
Interestingly, for the same connectivity parameter $m$, the
character time $t_c$  when the first maximum of the probabilities
occur on finite networks equals to that on infinite networks, i.e.,
the character time $t_c$ is independent on the size of the networks.

   The probabilities to go from a starting node to all other nodes in
time $t$ on a network of size $N=100$ with different values of $m$
are plotted in Fig.~\ref{fg3}. The starting excitation is located at
node $1$, and we can see the time propagating to the opposite node
$51$ becomes small on networks with large value of $m$. In addition,
the structure is quite regular when $m=1$. As $m$ increases, the
pattern becomes irregular.

\begin{figure}
\scalebox{0.8}[0.8]{\includegraphics{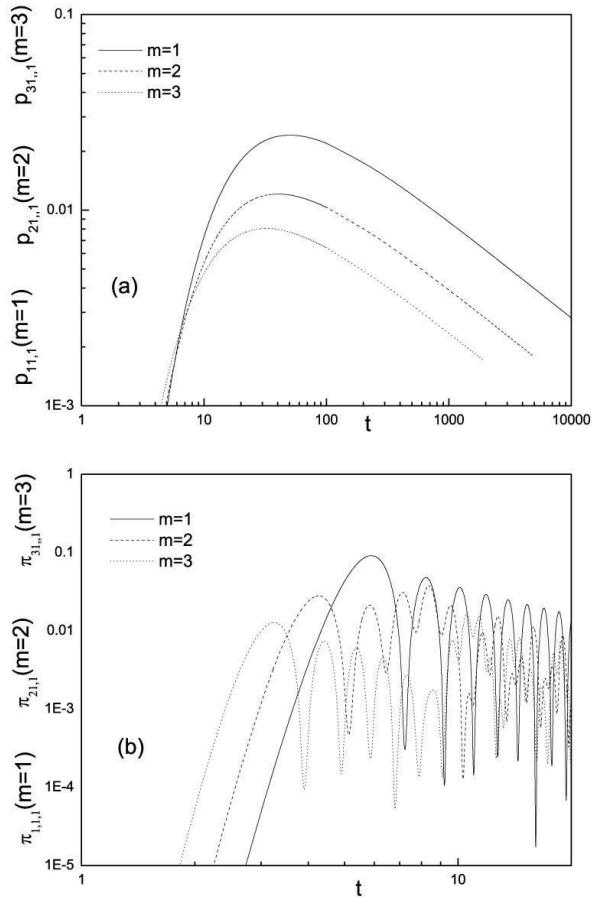}} \caption{ Time
evolution of transition probabilities on infinite networks for CTRWs
(a) and CTQWs (b). The initial excitation is located at node $1$.
The solid curves show the probabilities of being at node $11$ for
$m=1$ ($p_{11,1} $ in (a) and $\pi_{11,1}$ in (b)). Dashed curves
show the probabilities of being at node $21$ for $m=2$ ($p_{21,1} $
in (a) and $\pi_{21,1}$ in (b)). Dotted curves show the probability
of being at node $31$ for $m=3$ ($p_{31,1} $ in (a) and $\pi_{31,1}$
in (b)). The shortest path lengths between the two nodes are equal,
but the time when the first maximal value appears are different.
\label{fg4}}
\end{figure}

In order to compare the transport speed on different networks, we
define the character time $t_c$ as the time when the first maximum
of the probabilities occur on infinite networks. Such definition is
held both for the classical and quantum transport. For the classical
transport, there is only one maximal value and the character time
corresponds to the time when the equipartitioned probability $1/N$
is reached on finite networks. Now it is natural to ask the question
: Does the transport take equal time between two nodes of the same
shortest path length? To address this question, we calculate the
transition probabilities between two nodes having the same value of
shortest path length on infinite networks. Fig.~\ref{fg4} (a) shows
the classical transition probabilities $p_{11,1}$, $p_{21,1}$ and
$p_{31,1}$ for $m=1$, $m=2$ and $m=3$. The shortest path lengths of
the two nodes for the three infinite networks equal to $10$, but the
character time $t_c$ is small for highly connected networks. This
indicates that the transport is quick on networks with high
connectivity for CTRWs. For CTQWs, the same conclusion is also true,
as confirmed by the corresponding plot in Fig.~\ref{fg4} (b). The
character time $t_c$ for the quantum transport is much smaller than
that of the classical one, this supports the fact that the quantum
walks have efficient quantum algorithms with respect to their
classical counterparts~\cite{rn31}.

\begin{figure}
\scalebox{0.9}[0.9]{\includegraphics{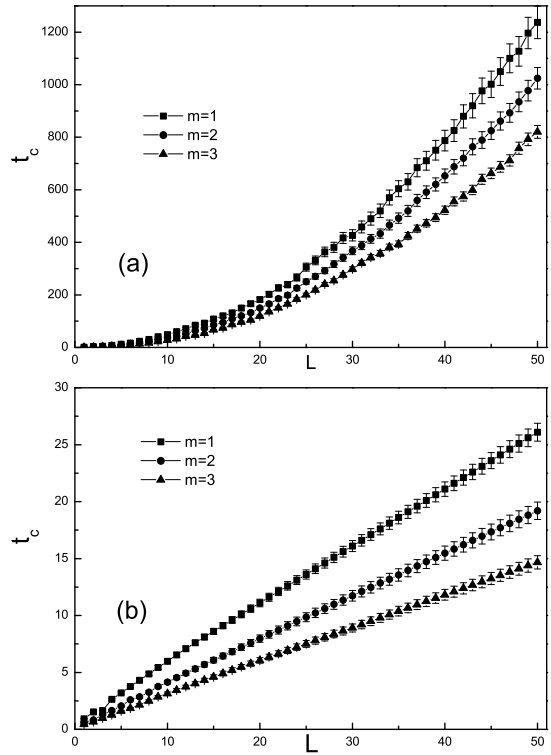}} \caption{ The
chacter time $t_c$ as a function of the shortest path length $L$
with different values of $m$ for CTRWs (a) and CTQWs (b). From the
figure, we can see that the classical transport gets slower while
the quantum transport velocity turns out to be invariable for a
certain value of $m$. \label{fg5} }
\end{figure}

 Fig. \ref{fg5} shows the character time $t_c$ versus the shortest path length
on networks with different values of $m$.  For classical transport
(Fig. \ref{fg5} (a)), $t_c$ grows faster than the shortest path
length $L$. It is found that the relationship between the character
time $t_c$ and the shortest path length $L$ can be well described by
quadratic equation: $t_c=\beta L^2$, where the parameter $\beta$ can
be obtained by fitting the data. Defining the transport speed $v$
 as the ratio of $L$ and $t_c$, we find that the classical transport speed gets slow for large $L$ while the
quantum transport speed turns out to be a constant values. We note
that the transport speed $v$ is large on highly connected networks
even the two nodes are located at the same distance $L_{i,j}$. By
fitting the linear relation between $t_c$ and $L$, the quantum
transport velocities are estimated to be about $1.92$, $2.62$ and
$3.41$ for $m=1$, $m=2$ and $m=3$ respectively. The different
behavior of the transport velocities between CTRQs and CTQWs is a
striking characteristic that distinguishes the classical and quantum
transport processes.
\section{Long time limiting probabilities}
Now, we consider the long time averaged probabilities. Classically,
the long time liming probabilities equal to the equipartitioned
probability $1/N$ \cite{rn21}. Quantum mechanically, the limiting
probabilities are determined by Eq. (\ref{eq18}) but the situation
is more complex for different network parameters. For $m=1$, the
spectral (or energy) of the system is $E_n=2-2\cos(\theta_n)$, where
$\theta_n=2n \pi /N$, $n\in [0,N)$. If the network size $N$ is an
even number, there are two nondegenerate eigenvalues, $E_{N/2}=4$
and $E_{0}=0$, and other eigenvalues have degeneracy 2. The limiting
probabilities can be written as
\begin{equation}\label{eq21}
 \chi_{ij}=\left\{
\begin{array}{ll}
2(N-1)/N^2,   & {\rm if} \ i=j \ , \ i=j\pm N/2,\\
(N-2)/N^2,    & Otherwise.
\end{array}
\right.
\end{equation}
    If the network size $N$ is an odd number, there are one
nondegenerate eigenvalues $E_{N}=0$, and the other eigenvalues have
degeneracy $2$. The limiting probabilities can be summarized as
\begin{equation}\label{eq22}
 \chi_{ij}=\left\{
\begin{array}{ll}
(2N-1)/N^2,   & {\rm if} \ i=j ,\\
(N-1)/N^2,    & Otherwise.
\end{array}
\right.
\end{equation}
Which confirms the results in Ref. \cite{rn28}.
\begin{figure}
\scalebox{0.9}[0.9]{\includegraphics{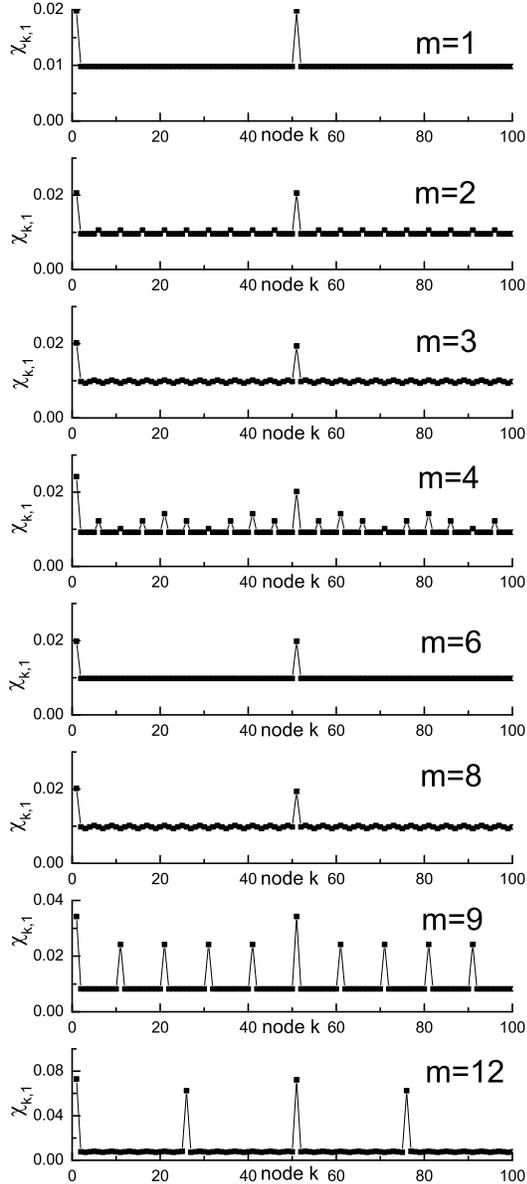}}
 \caption{(Color
online) The long time averaged probability distribution $\chi_{k,1}$
for CTQWs on networks of size $N=100$ with different values of $m$.
 \label{fg6}}
\end{figure}
\begin{figure}
\scalebox{0.9}[0.9]{\includegraphics{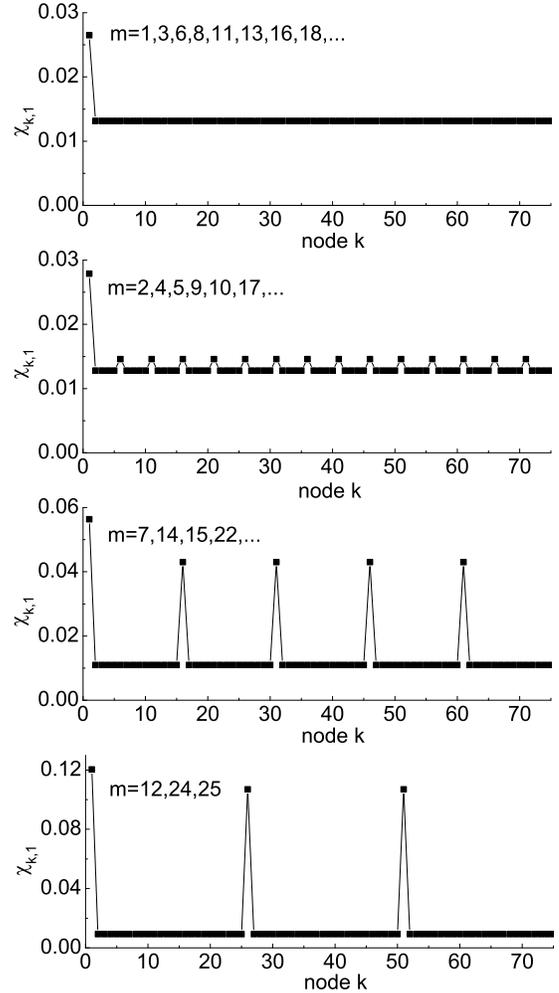}} \caption{(Color
online) Quantum mechanical limiting probabilities $\chi_{k,1}$ on
networks of size $N=75$ with different values of $m$.
 \label{fg7}}
\end{figure}

   For other values of $m$, the limiting probability distributions can also
be determined according to the degeneracy distribution of the
eigenvalues, but such process is complicated for large values of
$m$. Here, we report the limiting probabilities numerically obtained
using the Eq. (\ref{eq18}). In Fig. \ref{fg6}, we display the
limiting probabilities on the network of size $N=100$ with the
starting node $1$. As we can see, the probabilities for $m=6$ and
$m=8$ are the same as $m=1$ and $m=3$. After a careful examination,
we find that $m=8$ and $m=3$ have the same degeneracy distribution
of eigenvalues, and $m=6$ and $m=1$ have the same degenerate
eigenvalue distribution. Particularly, for all the values of $m$,
there is a large probability to be still or again at the initial
node and at the opposite node $k=51$. For some values of $m$, the
probabilities at the two positions are extremely high, for instance,
when $m=12$, the return probabilities exceed $0.07$.  For odd number
of network size, there is a higher probability to find the walker at
the initial node than that at other nodes. For networks of size
$N=101$ and $m\neq 50$, the limiting probability distribution shows
the same pattern described as Eq.(\ref{eq22}). One may conjecture
that the pattern of $\chi_{k,1}$ does not change when increasing
parameter $m$ on odd-numbered networks, but this is not true for
some values of network size $N$. For instance, on networks of size
$N=75$ with some particular values of $m$, the limiting probability
distribution differs from the pattern of Eq.(\ref{eq22}) (See
Fig.~\ref{fg7}). It is interesting to note that the patterns of
$\chi_{k,1}$ are the same for some values of $m$, this feature can
be explained by the identical degeneracy distribution of the
eigenvalues for different values of $m$.

\begin{figure}
\scalebox{0.9}[0.9]{\includegraphics{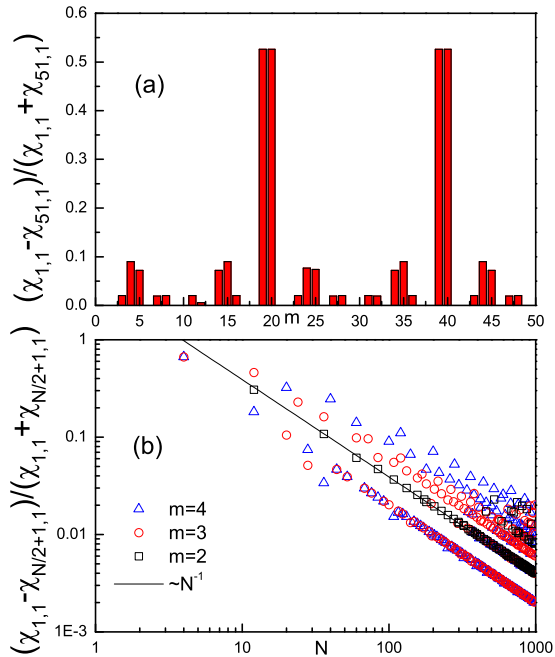}} \caption{(Color
online)(a) Relationship between the quantity $\Delta (1,50)\equiv
(\chi_{1,1}-\chi_{51,1})/(\chi_{1,1}+\chi_{51,1})$ and $m$ on a
network of size $N=100$. The nonzero value of $\Delta (1,50)$
presents asymmetry of probabilities $\pi_{1,1}$ and $\pi_{51,1}$.
(b)$\Delta(1,N/2)\equiv(\chi_{1,1}-\chi_{N/2+1,1})/(\chi_{1,1}+\chi_{N/2+1,1})$
versus network size $N$ for different values of $m$. The solid line
indicates the power law decay $\Delta(1,N/2)\sim N^{-1}$.
\label{fg8}}
\end{figure}

  As we have shown, if the network size $N$ is an even number, there
are high probabilities to find the walker at the initial node and
the opposite node. For some values of $m$, we find that the
probability of being at the initial node equals to the probability
of being at the opposite node. However, for some other values of
$m$, this is not true. In Ref. \cite{rn28}, the authors find
asymmetry of the probabilities for the starting node and its mirror
node, their definition of mirror node is based on geometry symmetry
of the network. In this paper, we define the mirror node $i^{'}$ of
a given node $i$ to be its opposite node, i.e., $i^{'}=i+N/2$. We
find asymmetry of the probabilities of being at the initial node and
at the opposite node (mirror node) for some particular network
parameters $N$ and $m$. Such asymmetry is small and not easy to be
observed in Fig.~\ref{fg6}. For a network of size $N=100$ and
assuming the initial exciton starts at node $1$, we find that
asymmetries occur at $m=3,4,5,7,8,11,12, 14,15,16, 19,20,$.... The
asymmetrical limiting probabilities are particularly characterized
by the difference between $\chi_{1,1}$ and $\chi_{51,1}$, thus we
use the quantity $\Delta (1,50)\equiv
(\chi_{1,1}-\chi_{51,1})/(\chi_{1,1}+\chi_{51,1})$ to detect the
asymmetry of the probabilities. In Fig.~\ref{fg8} (a), we present
$\Delta (1,50)$ as a function of parameter $m$. There are $29$
distinct values of $m$ having asymmetrical probabilities, which is
indicated by the nonzero value of $\Delta (1,50)$.

To reveal a general dependence of the asymmetry on the network
parameters, we plot the quantity $\Delta (1,N/2)\equiv
(\chi_{1,1}-\chi_{N/2+1,1})/(\chi_{1,1}+\chi_{N/2+1,1})$ as a
function of the network size $N$ for different values of $m$, which
are shown in Fig. \ref{fg8} (b). For $m=1$, the probabilities are
symmetrical for all the network size $N$, thus we only show the
asymmetry for $m=2$, $m=3$, and $m=4$. We find that
 the points break into several clusters, whereas some clusters $\Delta (1,N/2)$ decreases with
the network size $N$ as a power law: $\Delta (1,N/2)\sim N^{-1}$.

Except for the asymmetrical probabilities between the initial node
and the opposite node (mirror node), we also find asymmetrical
probabilities between other nodes and their mirror nodes. In our
calculations, we find that such asymmetries can be different from
the asymmetry of the probability of being at the initial node and
being at its opposite node. For instance, considering a CTQW on a
network of size $N=100$ and assuming the initial excitation starts
at node $1$, there are asymmetries between $\chi_{1+n,1}$ and
$\chi_{51+n,1}$ ($n$ is an even number) for some values of $m$. The
discrete values of $m$ for different asymmetries can differ from
each other, depending on the precise value of $N$ and $m$. This
situation is even more complex and requires a further study.

\section{Conclusions and Discussions}
In summary, we have studied continuous-time quantum walks on
one-dimension ring lattice of $N$ nodes in which each node is
connected to its $2m$ nearest neighbors ($m$ on either side).  Using
the Bloch function approach, we calculate transition probabilities
between two nodes of the lattice, and compare the results with the
classical counterpart. It is found that the transport of CTQW is
faster than that of the classical continuous-time random walk. We
define the transport velocity as the ratio of the shortest path
length and spreading time between two nodes. For network of a given
parameter $m$, the transport of CTRWs gets slow with the increase of
the shortest distance while the transport of CTQWs spreads the
network constantly. In the long time limit, depending on the network
parameters $N$ and $m$, the limiting probability distributions of
CTQWs show various patterns. When the network size $N$ is an even
number, the probability of being at the original node differs from
that of being at the opposite node, which also depends on the
precise value of parameter $m$. Asymmetrical probabilities between
other nodes and their mirror nodes also exist for some particular
network parameters.

    The asymmetry of the limiting probabilities of being at a node and being at its mirror
node is a novel phenomenon, which does not exist in the cycle graph
with $m=1$. However, we are unable to predict which particular
parameters of $N$ and $m$ are related to such asymmetry.
Furthermore, in our calculations, we find a large value of the
limiting return probability for some special network topology, for
instance, on a complete graph in which each pair of nodes is
connected, the long time averaged return probabilities equal to
$\chi_{j,j}=(N^2-2N+2)/N^2$ while the other transition probabilities
are $\chi_{k,j}=2/N^2$ ($k\neq j$). This is a striking feature of
CTQWs which differs from the classical counterpart.

\begin{acknowledgments}
The authors would like to thank Zhu Kai for converting the
mathematical package used in the calculations. This work is
supported by the Cai Xu Foundation for Research and Creation (CFRC),
National Natural Science Foundation of China under projects
10575042, 10775058 and MOE of China under contract number IRT0624
(CCNU).
\end{acknowledgments}


\begin{thebibliography} {Albert2000}
\bibitem{rn32} T. Odagaki and M. Lax, Phys. Rev. B {\bf 24}, 5284 (1981).
\bibitem{rn33} T. Odagaki and M. Lax, Phys. Rev. B {\bf 26}, 6480 (1982).
\bibitem{rn1} G. H. Weiss, \emph{Aspect and Applications of the Random Walk} (North-Holland, Amsterdam, 1994).
\bibitem{rn2} J. Kempe, Contemp. Phys. {\bf 44}, 307 (2002).
\bibitem{rn3} D. Supriyo, \emph{Quantum Transport: Atom to Transistor} (Cambridge University Press, London, 2005).
\bibitem{rn4} A. Ambainis, \emph{Quantum search algorithms} (New York, USA , 2004).
\bibitem{rn5} Y. Aharonov, L. Davidovich, and N. Zagury, Phys. Rev. A {\bf 48}, 1687 (1993).
\bibitem{rn6} N. Shenvi, J. Kempe, and K. Brigitta Whaley, Phys. Rev. A {\bf 67}, 052307 (2003).
\bibitem{rn7} E. Farhi and S. Gutmann, Phys. Rev. A {\bf 58}, 915 (1998).
\bibitem{rn8} A. M. Childs, E. Farhi, and S. Gutmann, Quant. Inf. Proc. {\bf 1}, 35 (2002).
\bibitem{rn9} H. Gerhardt and J. Watrous, quant-ph/0305182.
\bibitem{rn10} N. Konno, Phys. Rev. E {\bf 72}, 026113 (2005).
\bibitem{rn11} A. Ambainis, J. Kempe  and A. Rivosh,\emph{Coins make quantum walks faster} (Philadelphia, USA, 2005).
\bibitem{rn12} D. Shapira, O. Biham, A. J. Bracken, and M. Hackett, Phys. Rev. A {\bf 68}, 062315 (2003)
\bibitem{rn13} N. Ashwin and V. Ashvin, quant-ph/0010117.
\bibitem{rn14} G. Abal, R. Siri, A. Romanelli, et al., Phys. Rev. A {\bf 73}, 042302 (2006)
\bibitem{rn15} D. Solenov and L. Fedichkin, Phys. Rev. A {\bf 73}, 012313 (2003)
\bibitem{rn16} F. Sorrentino, M. di Bernardo, G. H. Cu\'ellar, and S. Boccaletti, Physica D {\bf 224}, 123 (2006).
\bibitem{rn17} H. Krovi and T. A. Brun, Phys. Rev. A {\bf 73}, 032341 (2006).
\bibitem{rn18} O. M\"{u}lken and A. Blumen, Phys. Rev. E {\bf 71}, 016101 (2005).
\bibitem{rn19} O. M\"{u}lken, V. Bierbaum and A. Blumen, J. Chem. Phys {\bf 124}, 124905 (2006).
\bibitem{rn20} O. M\"{u}lken and A. Blumen, Phys. Rev. E {\bf 71}, 036128 (2005).
\bibitem{rn21} A. Volta, O. M\"{u}lken and A. Blumen, J. Phys. A {\bf 39}, 14997 (2006).
\bibitem{rn22} S. H. Strogatz, and I. Stewart, Sci. Am. {\bf 269}, 102 (1993).
\bibitem{rn23} K. Wiesenfeld, Physica B {\bf 222}, 315 (1996).
\bibitem{rn24} L. F. Abbott and C. V. Vreeswijk, Phys. Rev. E {\bf 48}, 1483 (1993).
\bibitem{rn25} I.V. Belykh, V.N. Belykh and M. Hasler, Physica D {\bf 195}, 159 (2004).
\bibitem{rn26} D. J. Watts and S. H. Strogatz, Nature {\bf 393}, 440 (1998).
\bibitem{rn27} C. Kittel, \emph{Introduction to solid state physics} (Wiley, New York, 1986).
\bibitem{rn28} O. M\"{u}lken, A. Volta and A. Blumen, Phys. Rev. A {\bf 72}, 042334 (2005)
\bibitem{rn29} G. E. Andrews, R. Askey and R. Roy, \emph{Special functions} (Tsinghua Univ, China, 2004).
\bibitem{rn30} O. M\"{u}lken and A. Blumen, Phys. Rev. E {\bf 73}, 066117 (2006).
\bibitem{rn31} P. C. Richter, Phys. Rev. A {\bf 76}, 042306 (2007).

\end{thebibliography}
\end{document}